# Distinguishing fissions of $^{232}$Th, $^{237}$Np and $^{238}$U with beta-delayed gamma rays

A. Iyengar[1], E.B. Norman [1], C. Howard [1], C. Angell [1], A. Kaplan [1], J. J. Ressler[2], P. Chodash[1], E. Swanberg[1], A. Czeszumska[1], B. Wang[1], R. Yee[1], H.A. Shugart[3]

[1]*Department of Nuclear Engineering, University of California, Berkeley, CA 94720, USA*
[2]*Lawrence Livermore National Laboratory, Livermore, CA 94551, USA*
[3]*Department of Physics, University of California, Berkeley, CA 94720, USA*

**ABSTRACT**

Measurements of beta-delayed gamma-ray spectra following 14-MeV neutron-induced fissions of $^{232}$Th, $^{238}$U, and $^{237}$Np were conducted at Lawrence Berkeley National Laboratory's 88-Inch Cyclotron. Spectra were collected for times ranging from 1 minute to 14 hours after irradiation. Intensity ratios of gamma-ray lines were extracted from the data that allow identification of the fissioning isotope.

# 1. Introduction

When a nucleus undergoes fission, the products emit both prompt and delayed characteristic gamma-rays. The delayed gamma-rays are emitted following beta-decays, and therefore also have characteristic time scales associated with their emissions. These gamma-rays can be used to identify fissionable material hidden in cargo containers, or distinguish between fissionable isotopes, and are useful for assay of nuclear fuel. They can also be potentially used in nuclear forensics to identify the average energy of neutrons producing fission and the original fissionable isotope. Previous works measuring beta-delayed gamma-rays from neutron-induced fission focused on $^{238}$U and the major fissile isotopes, $^{235}$U and $^{239}$Pu. [1], [2], [3], [4].

However, for beta-delayed gamma-ray spectroscopy to be a useful tool, a complete set of measurements on all likely fissionable isotopes is needed. Measurement of beta-delayed gamma-rays from fissions of $^{232}$Th, and $^{237}$Np will help expand the database, making it a viable tool in detection, assay, and nuclear safeguards applications. The fast neutron induced fissions of $^{238}$U, $^{232}$Th and $^{237}$Np produce complex fission spectra for many hours following irradiation, samples of $^{238}$U, $^{232}$Th and $^{237}$Np were irradiated with 14-MeV neutrons in order to measure the beta-delayed gamma-ray spectra from induced fission. We used both high-resolution germanium detectors and low-resolution plastic scintillators to measure gamma energy and time spectra. Spectra from both sets of detectors from 1 minute to 14 hours after irradiation were found to contain distinguishing features that can be used to identify the fissioning isotope. We present distinctive gamma-ray intensity ratios for each sample in several time bins, which provide adequate data to identify the fissionable material present.

# 2. Experimental method

Samples of ~ 2 g each of $^{238}$U, $^{232}$Th and $^{237}$Np oxides obtained from UC Berkeley and Lawrence Berkeley National Laboratory were irradiated at the 88-Inch Cyclotron at Lawrence Berkeley National Laboratory with neutrons produced from a deuteron beam incident on a thick beryllium target. A 33-MeV, 1.0-µA deuteron beam was used to produce a neutron energy distribution centered at 14 MeV. [5]. Each of the oxide samples was doubly encapsulated in polyethylene capsules. The fissionable samples were placed immediately behind the beryllium neutron-production target. The samples were covered with 1 mm of cadmium to exclude thermal neutrons.

Each sample was exposed to neutrons for 1 minute, and samples were removed immediately after exposure and counted with gamma-ray detectors for a long sequence of time intervals beginning approximately 1 minute after the end of neutron irradiation. The samples were quickly moved from the irradiation location and placed equidistant from a 10x10x15-cm plastic scintillator and a 65% relative efficiency high-purity germanium detector for 10 one-minute counting intervals. The samples were placed 5 cm from each detector face. Approximately 2 cm of lead and 3 cm of polyethylene were placed between the sample and the detectors in order to attenuate electrons and low-energy gamma rays. Separate well-shielded HPGe detectors were then used to continue counting each sample for approximately 14 hours. Combinations of lead and plastic were also used in these measurements to attenuate betas and low energy gammas. All detectors were energy-calibrated with $^{137}$Cs, $^{22}$Na and $^{232}$Th sources. After counting of the all the targets was completed, careful gamma-ray efficiency measurements were performed with a variety of calibrated sources covering the energy range up to 1.4 MeV.

*2.1 Time Bins*

Following irradiation, spectra were recorded for 10 1-minute time bins, followed by 15 3-minute time bins, followed by a variable number of 10-minute time bins. This enabled the observation of all significant spectral changes with time scales ranging from minutes to approximately half a day. To simplify the analysis and reporting of the measurements, the experimental data was combined into six larger time bins: 5–8 m, 18–33 m, 1–1.5 h, 2.5–3.5 h, 5–7 h, and 10–14 h after the irradiation period. These time bins are long enough to obtain good statistical precision, and they span the full time range of our data set. These same time intervals were selected by Marrs et. al [1]. for their complete coverage of the data set and their capture of all important time-dependent features in the fission-product spectra such that there are no major gaps in temporal coverage. In what follows, we report experimental line ratios for these six time intervals; however, a different choice of time intervals would yield similar results and the same conclusions.

*2.2 Peak Identification and Peak-Area Extraction*

Peak identification and area extraction on the energy spectra from the HPGe detector was performed manually using the Ortec Maestro software. The extracted peak areas were then corrected for the energy-dependent detection efficiency before intensity ratios were determined. Figures 1(a) and 1(b) illustrate summed gamma-ray spectra from the germanium and plastic scintillator detectors, respectively, for the first 10 minutes of counting the $^{237}$Np Sample. From Figure 1(a) it can be seen that the Germanium spectrum is extremely complicated due to the presence of numerous fission products and their many gamma-ray lines.

*2.3 Line Pair Selection*

Line pairs selection criteria were specified by Marrs et. al. The basic criterion to be fulfilled is that the intensity ratios at corresponding energies need to differ significantly between $^{232}$Th, $^{237}$Np, and $^{238}$U. Line pairs also needed to be close in energy to minimize the energy dependent corrections for absorption and detector efficiency. Peaks below 500 keV were avoided due to the complexity of the spectra and high background in that energy range. Gamma rays selected as line pair candidates were validated whenever possible by identifying other lines from those isotopes and comparing measured intensities to those predicted pre-experiment using online nuclear databases [6, 7, 8].

## 3. Fissile Material Identification

As stated before, the goal was to look for line pairs with differing intensity ratios between the isotopes of interest. Table 1 shows the gamma ray ratios for different time ranges comparing $^{232}$Th, $^{237}$Np and $^{238}$U. Notice that there was no single ratio that could be used to identify all three isotopes within the 5 - 8 minute time bin. Figures 2-6 show a visual comparison of all three lines plotted at the peak energies of interest.

## 4. Scintillator Data Analysis

Ten one minute spectra were collected with both, an HPGe detector and a plastic scintillator detector immediately after irradiation of each sample. The spectra from the scintillator do not have distinct peaks due to weaker energy resolution compared to the HPGe as can be seen in Figures 1(a) and 1(b). However, these detectors provide valuable data that demonstrate different trends over time and energy for different fissile materials. By integrating over certain energy ranges, we can compare the trends and ratios in the total counts seen in each sample in order to distinguish between the materials. The integrated counts can also be measured as a function of time to determine decay constants specific for each material.

Energies below 3 MeV are not considered in analysis to exclude any natural background. The count rate drops significantly above approximately 6.5 MeV and after 6 or 7 minutes, resulting in very high statistical uncertainty. For these reasons, the optimal time and energy range for material identification is 1-6 minutes following irradiation and 3-6.5 MeV. Table 2 shows ratios of counts integrated over 3-3.5 MeV versus 5-5.5 MeV, 6-6.5 MeV, and 7-7.5 MeV ranges for $^{237}$Np, $^{232}$Th and $^{238}$U.

From this table, it can be seen that at higher energies, the ability to distinguish between the three isotopes improves. Also, ratios of $^{237}$Np and $^{238}$U stay fairly close in value, and the $^{232}$Th values differ from the $^{237}$Np and $^{238}$U values more. From the summed gamma-ray spectra, it is verified that it is difficult to differentiate $^{238}$U from $^{237}$Np from scintillator spectra alone. However, it is easy to distinguish $^{232}$Th from $^{238}$U and $^{237}$Np.

## 5. Conclusions

We have shown that there are several different ways to distinguish between the fast fission of $^{237}$Np, $^{232}$Th and $^{238}$U. When these materials fission, a broad range of fission products are created all emitting beta delayed gamma-rays at characteristic time scales. Different fissile materials produce fission products in different abundances. By examining spectra at different times following induced fission, it is possible to identify the most prominent of these gamma-rays and use them to distinguish between the fissile materials.

The best line pairs for differentiating materials have been identified for several different time bins, ranging from 5 minutes to 14 hours. Fission spectra measured by a plastic scintillator have also proven to be useful in identifying fissile material by comparing, quantitatively, the shape of the spectra, the decrease in counts seen over time, and the different behavior of the materials in specific energy bins.

From the $^{237}$Np, $^{232}$Th and $^{238}$U fission product spectra, it was possible to identify common line pairs to help distinguish them individually. Our results show it is relatively easy to distinguish $^{232}$Th from $^{237}$Np and $^{238}$U however, it is difficult to distinguish $^{237}$Np from $^{238}$U. Some of the experimental line pair ratios are much different from the theoretical ratios derived [6]. Somewhat surprisingly, the discrepancies between the theoretical ratios and those measured here appear larger for the 14-MeV neutron induced fission of $^{238}$U than for those of $^{232}$Th or $^{237}$Np. Several large discrepancies of this kind were also noted in the work of Marrs *et al.* [1]. Possible explanations for these differences between theory and experiment are that some fission yields in the England and Rider compilations may be incorrect, or that some fission product beta-delayed gamma-ray intensities may be incorrect. Dedicated high-precision measurements of both fission yields and fission product beta decay schemes would be needed to resolve these issues.

## 6. Acknowledgements

We wish to thank the 88-Inch Cyclotron operations and facilities staff for their help in performing this experiment. This work was supported in part at Lawrence Berkeley National Laboratory by the Director, Office of Energy Research, Office of High Energy and Nuclear Physics, Division of Nuclear Physics, of the U.S. Department of Energy under Contract DE-AC02-05CH11231; the U.S. Department of Energy, Lawrence Livermore National Laboratory under Contract DE-AC52-07NA27344; the Department of Energy, NNSA, Office of Non-Proliferation (NA-22); the U.S. Dept. of Homeland Security under contract number ARI-022, and NSERC.

| Time Bin | Line Pair | Fission Isotope | Experimental Intensity Ratio | Theoretical Intensity Ratio |
|---|---|---|---|---|
| 5-8 m | $\dfrac{^{89}\text{Rb (1032)}}{^{132}\text{Sb (974)}}$ | $^{238}$U | 0.30 ± 0.020 | 0.22 |
|  |  | $^{232}$Th | 1.03 ± 0.050 | 1.97 |
| 5-8 m | $\dfrac{^{89}Rb\ (832)}{^{132m}Sb\ (794)}$ | $^{232}$Th | 3.023 ± 0.450 | 3.206 |
|  |  | $^{237}$Np | 0.896 ± 0.050 | 0.448 |
| 18-33 m | $\dfrac{^{89}Rb\ (1032)}{^{138}Cs\ (1010)}$ | $^{232}$Th | 3.270 ± 0.310 | 2.821 |
|  |  | $^{237}$Np | 1.382 ± 0.050 | 1.032 |
|  |  | $^{238}$U | 2.260 ± 0.050 | 1.400 |
|  | $\dfrac{^{138}Cs\ (2218)}{^{89}Rb\ (2196)}$ | $^{232}$Th | 0.572 ± 0.055 | 0.876 |
|  |  | $^{237}$Np | 1.862 ± 0.103 | 2.150 |
|  |  | $^{238}$U | 0.877 ± 0.040 | 1.470 |
| 1-1.5 h | $\dfrac{^{89}Rb\ (1032)}{^{134}I\ (1072)}$ | $^{232}$Th | 1.893 ± 0.246 | 1.806 |
|  |  | $^{237}$Np | 0.591 ± 0.047 | 0.511 |
|  |  | $^{238}$U | 0.500 ± 0.020 | 0.260 |
|  | $\dfrac{^{89}Rb\ (2196)}{^{132}Cs\ (2218)}$ | $^{232}$Th | 0.433 ± 0.056 | 0.416 |
|  |  | $^{237}$Np | 0.240 ± 0.017 | 0.152 |
|  |  | $^{238}$U | 0.300 ± 0.020 | 1.200 |
| 2.5-3.5 h | $\dfrac{^{92}Sr\ (1384)}{^{135}I\ (1261)}$ | $^{232}$Th | 6.080 ± 0.730 | 6.036 |
|  |  | $^{237}$Np | 3.988 ± 0.144 | 3.136 |
|  |  | $^{238}$U | 3.530 ± 0.09 | 5.540 |
| 5-7 h | $\dfrac{^{92}Sr\ (1384)}{^{135}I\ (1261)}$ | $^{232}$Th | 3.636 ± 0.298 | 3.872 |
|  |  | $^{237}$Np | 2.207 ± 0.064 | 2.011 |
|  |  | $^{238}$U | 2.060 ± 0.040 | 3.570 |
| 10-14 h | $\dfrac{^{92}Sr\ (1384)}{^{135}I\ (1261)}$ | $^{232}$Th | 2.099 ± 0.159 | 1.614 |
|  |  | $^{237}$Np | 0.980 ± 0.035 | 0.839 |
|  |  | $^{238}$U | 0.790 ± 0.020 | 1.490 |

**Table 1:** Best intensity ratios for the fission products of $^{237}$Np, $^{232}$Th and $^{238}$U. Line pair ratios to identify all three were found and are shown for all time bins except the 5-8 min interval. It was necessary to use different line pairs to identify all three targets in the 5-8 min time bin.

| Ratio | $^{232}$Th | $^{238}$U | $^{237}$Np |
|---|---|---|---|
| N(3-3.5 MeV)/N(5-5.5 MeV) | 4.63 | 6.02 | 5.95 |
| N(3-3.5 MeV)/N(6-6.5 MeV) | 10.1 | 14.4 | 15.1 |
| N(3-3.5 MeV)/N(7-7.5 MeV) | 27.5 | 62.5 | 49.5 |

**Table 2:** Ratios of number of counts observed in the plastic scintillator integrated over different time bins for each of the target isotopes. Note that no corrections for energy-dependent gamma ray detection efficiencies have been applied to any of these data.

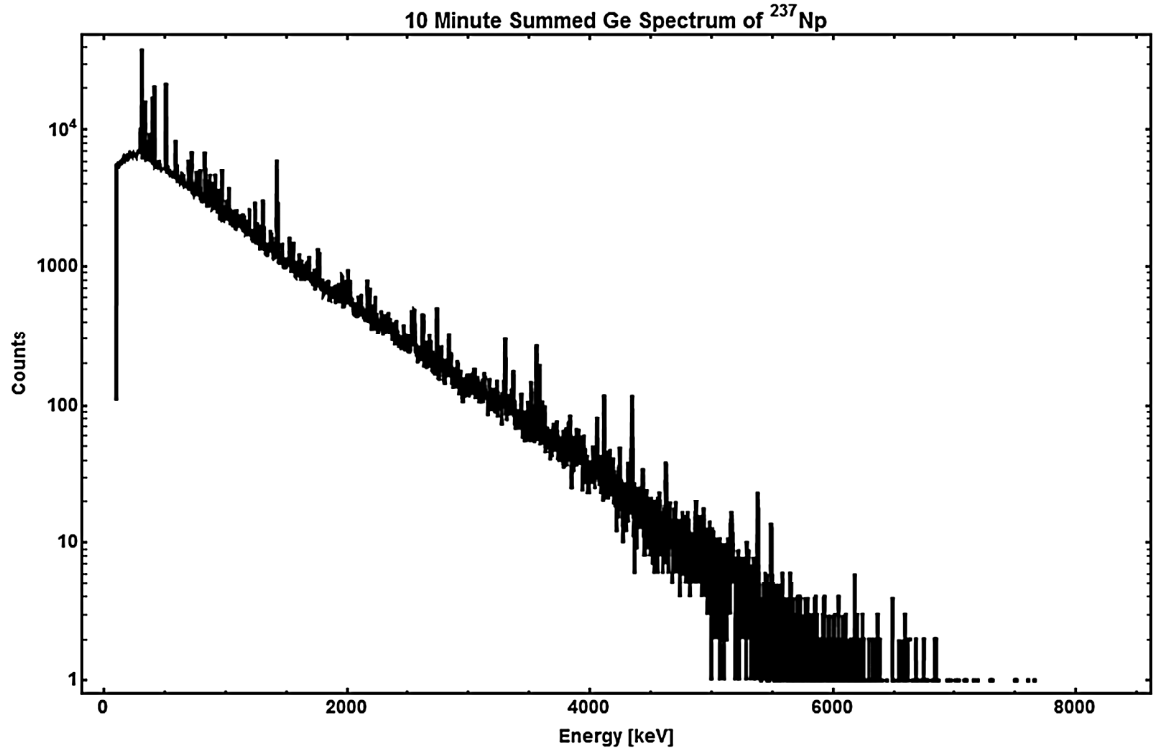

**Fig 1(a):** HPGe gamma-ray spectrum of $^{237}$Np fission products from the first 10 minutes of counting after irradiation.

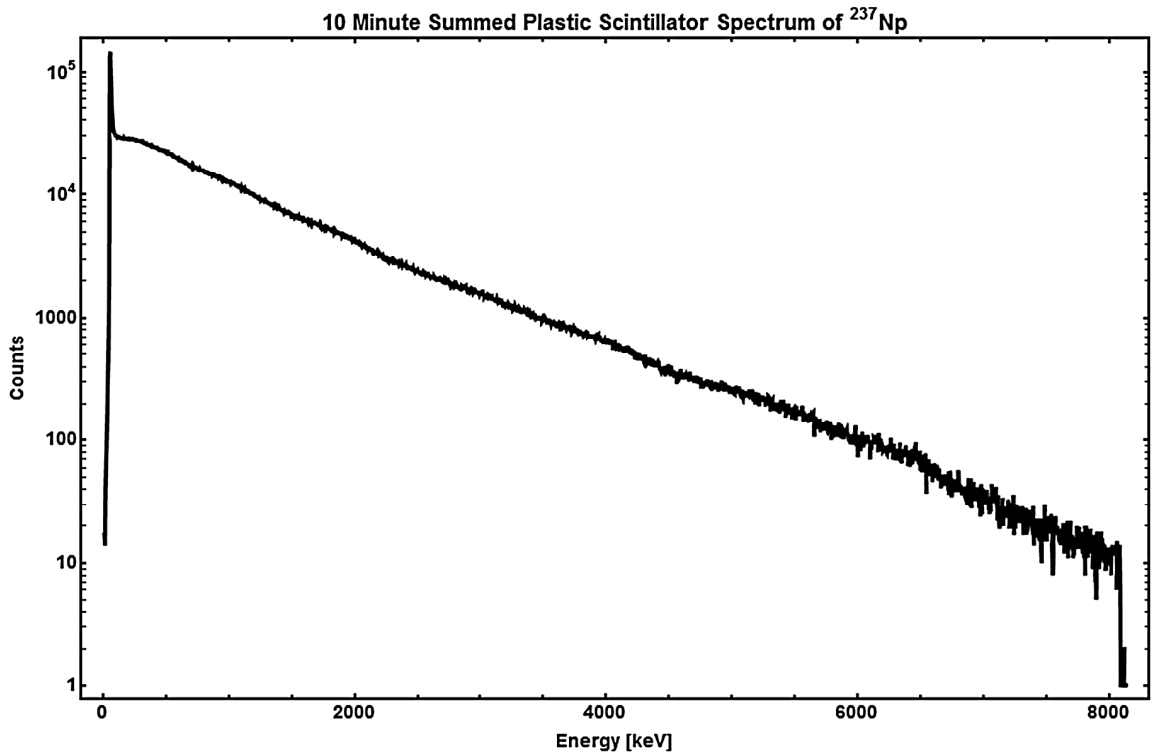

**Fig 1(b):** Plastic scintillator gamma-ray spectrum of $^{237}$Np fission products from the first 10 minutes of counting after irradiation.

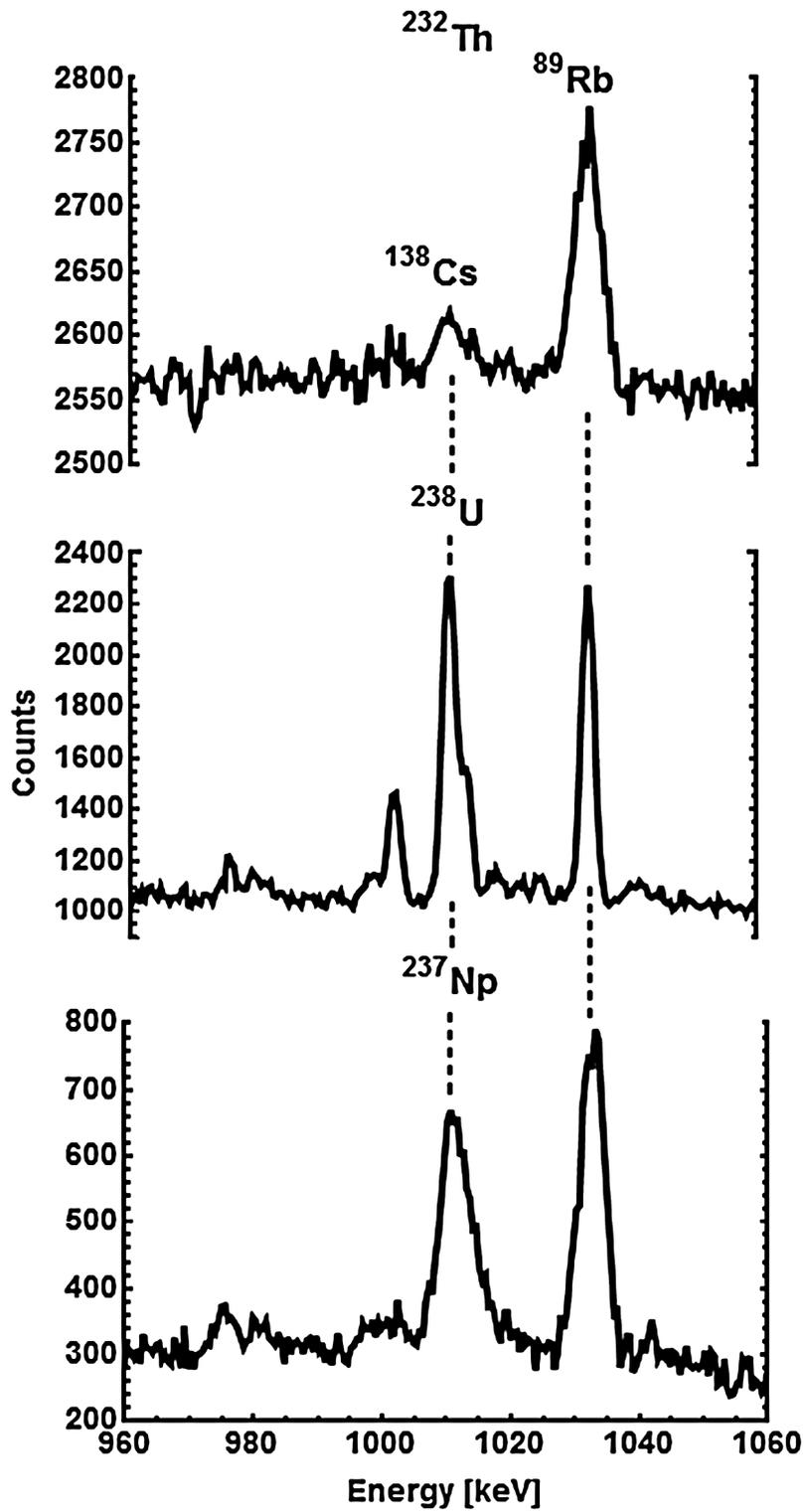

**Fig. 2:** Comparison of gamma-ray spectra from $^{232}$Th, $^{238}$U, and $^{237}$Np fission by 14 MeV peak neutrons, integrated from 18 to 33 min after fission. Line pairs shown here are from $^{138}$Cs (1010 keV) and $^{89}$Rb (1032 keV) peaks.

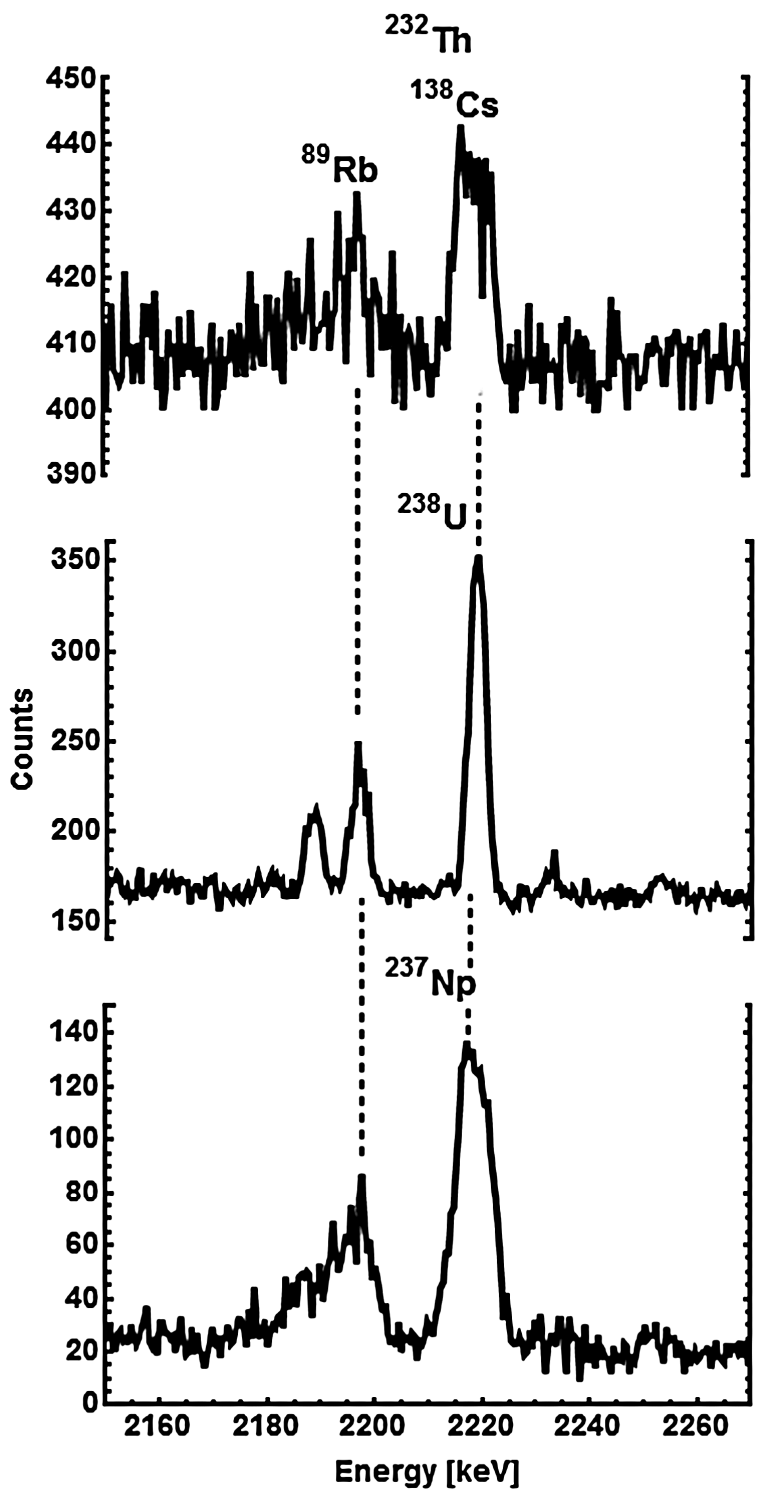

**Fig. 3:** Comparison of gamma-ray spectra from $^{232}$Th, $^{238}$U, and $^{237}$Np fission by 14 MeV peak neutrons, integrated from 1 to 1.5 hours after fission. Line pairs shown here are from $^{89}$Rb (2196 keV) and $^{134}$I (2218 keV) peaks.

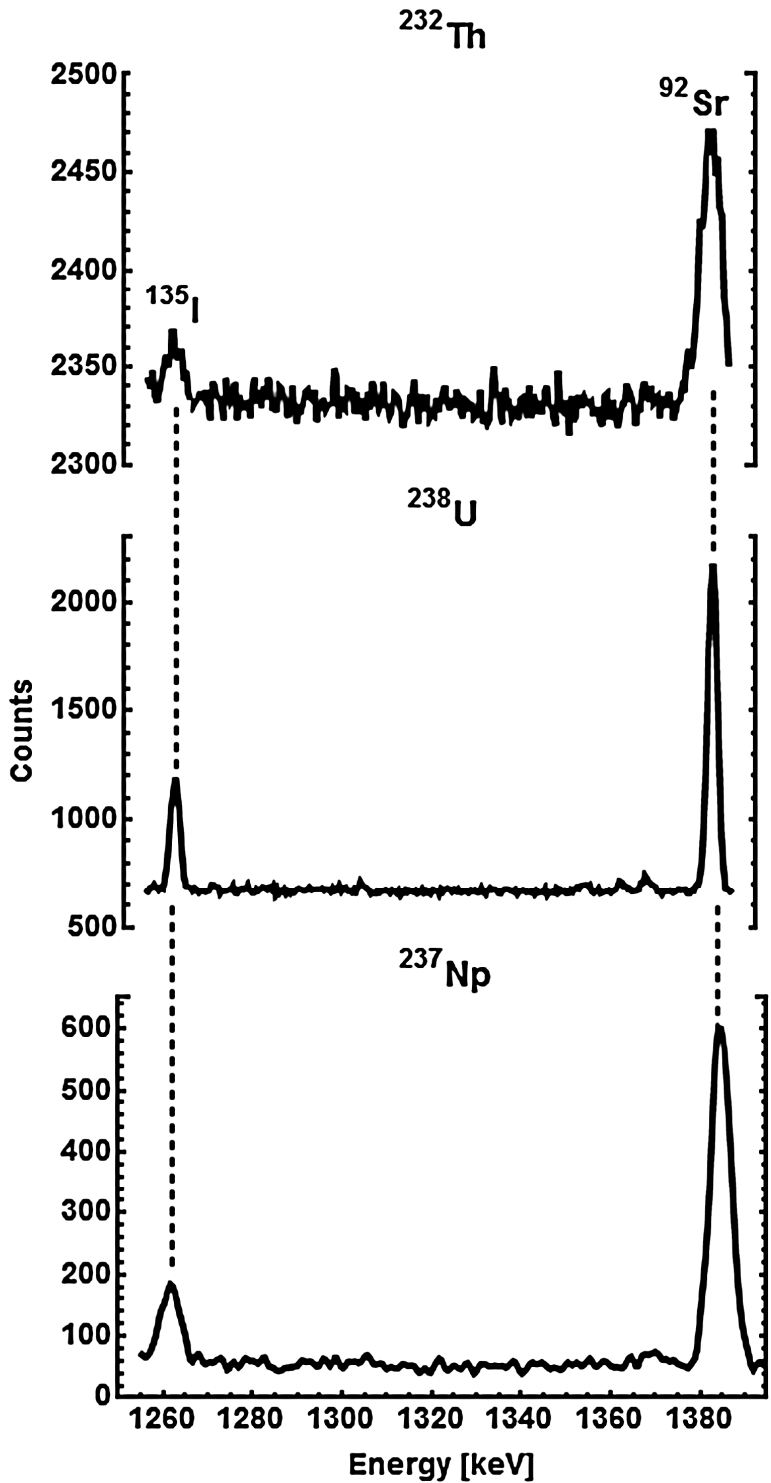

**Fig. 4:** Comparison of gamma-ray spectra from $^{232}$Th, $^{238}$U, and $^{237}$Np fission by 14 MeV peak neutrons, integrated from 2.5 to 3.5 hours after fission. Line pairs shown here are from $^{135}$I (1384 keV) and $^{92}$Sr (1260 keV) peaks.

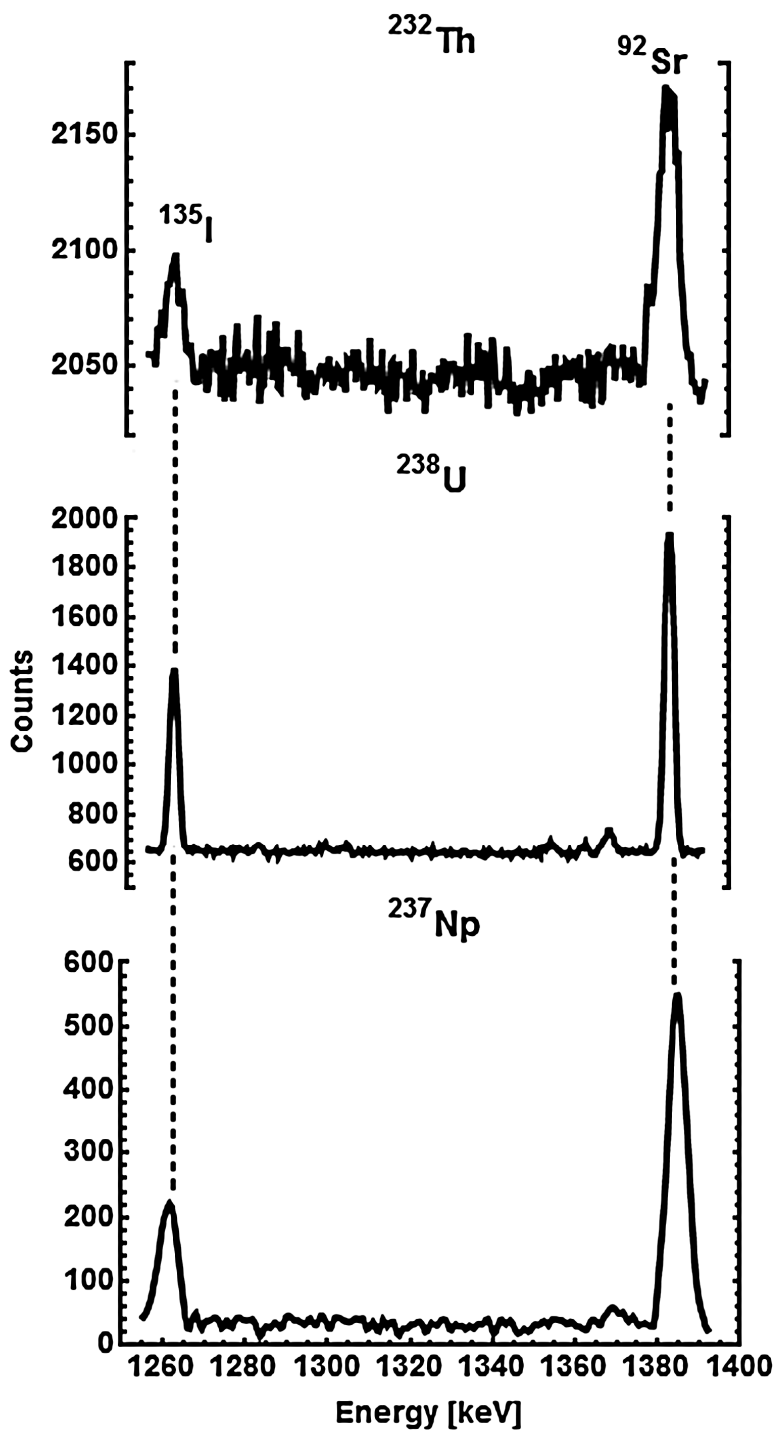

**Fig. 5:** Comparison of gamma-ray spectra from ²³²Th, ²³⁸U, and ²³⁷Np fission by 14 MeV peak neutrons, integrated from 5 to 7 hours after fission. Line pairs shown here are from ¹³⁵I (1384 keV) and ⁹²Sr (1260 keV) peaks.

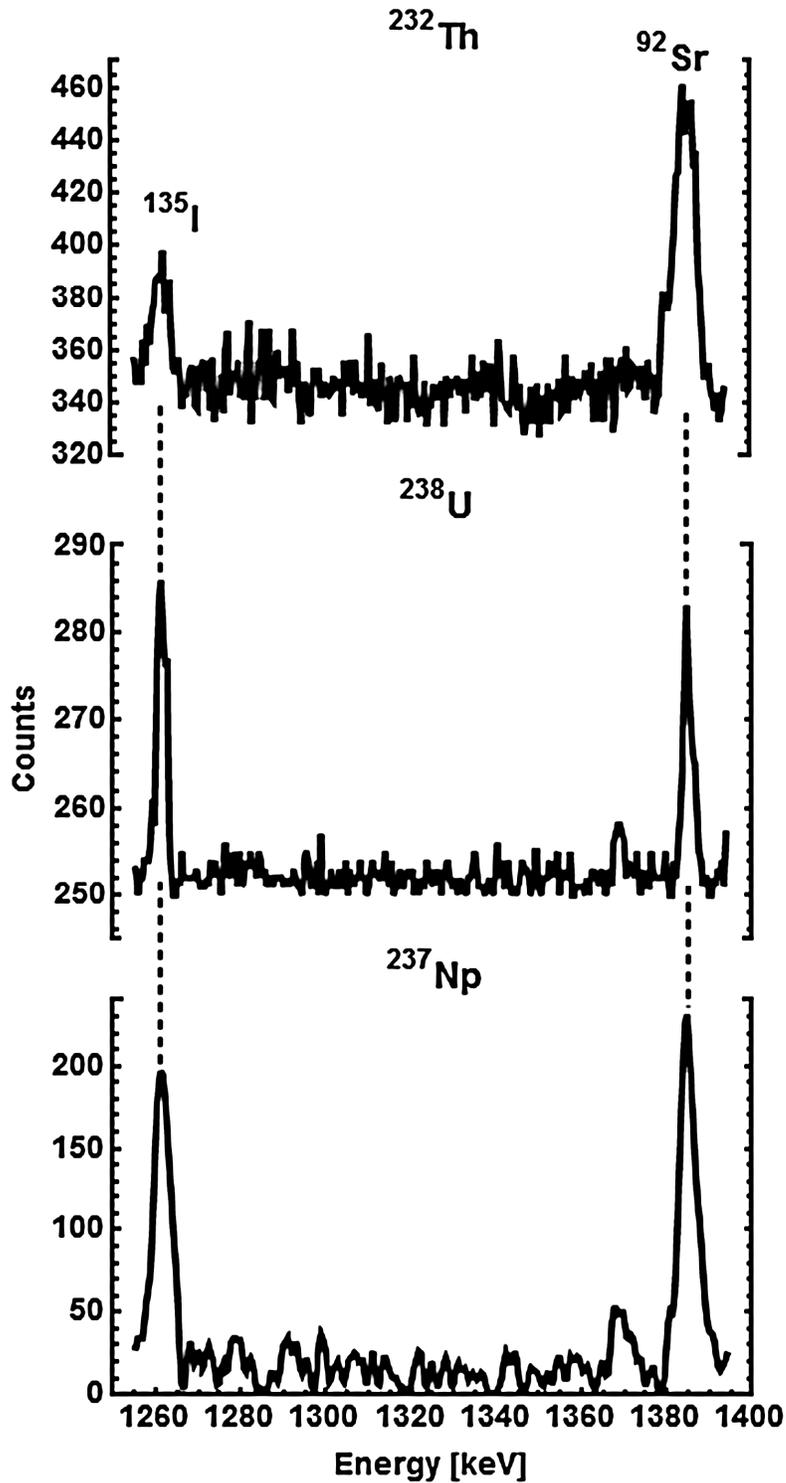

**Fig. 6:** Comparison of gamma-ray spectra from $^{232}$Th, $^{238}$U, and $^{237}$Np fission by 14 MeV peak neutrons, integrated from 10 to 14 hours after fission. Line pairs shown here are from $^{135}$I (1384 keV) and $^{92}$Sr (1260 keV) peaks.